\documentclass[aps,groupedaddress,showpacs,floatfix]{revtex4}
\usepackage{amssymb}
\usepackage{amsmath}
\usepackage{graphicx}
\begin{document}

\title{One more discussion of the replica trick: 
       the examples of exact solutions}

\author{Victor Dotsenko}

\affiliation{LPTMC, Universit\'e Paris VI, 75252 Paris, France}

\affiliation{L.D.\ Landau Institute for Theoretical Physics,
   119334 Moscow, Russia}

\date{\today}

\begin{abstract}
A systematic replica field theory calculations are analysed using the examples
of two particular one-dimensional "toy" random models with Gaussian disorder. 
Due to apparent simplicity of the model the replica trick calculations
can be followed here step by step from the very beginning till the very end. 
In this way it can be easily demonstrated that {\it formally} at certain stage of the
calculations the implementation of the standard replica program is just impossible. 
On the other hand, following the usual "doublethink"  traditions  of the replica 
calculations  (i.e. closing eyes on the fact that certain suggestions used 
in the calculations contradict to each other) one can easily fulfil the programme 
till the very end to obtain  physically sensible result for the entire free energy 
distribution function.

\end{abstract}


\maketitle

\section{Introduction}

In recent years there is a renewed  interest to the mathematical status of the 
replica method widely used in disordered systems during last four decades.
For the calculation of thermodynamic quantities averaged over disorder parameters
(e.g. average free energy) the method assumes, first, calculation of the averages
of an integer $n$-th power of the partition function $Z(n)$, and second, analytic continuation of 
this function in the replica parameter $n$ from integer to arbitrary non-integer values
(and in particular, taking the limit $n\to 0$). Usually one is facing difficulties at both
stages of this program. First of all, in  realistic disordered systems
the calculations of the replica partition function $Z(n)$ can be done only using
some kind of approximations, and in this case the status of further analytic continuation 
in the replica parameter $n$ becomes rather indefinite since the terms neglected 
at integer $n$ could become  essential at non-integer $n$ (in particular the limit $n\to 0$)
\cite{zirnbauer1,zirnbauer2}. 
The typical example of such type of trouble is provided by the classical Kardar's solution 
of $(1+1)$ directed polymers
in random potential where due to the approximation used at the first stage of calculations
(when the parameter $n$ is still integer)
the resulting free energy distribution function appears to be  not positively defined
\cite{kardar1,kardar2} (see also \cite{dirpoly}).
On the other hand, even in rare  cases when 
the derivation of the replica partition function $Z(n)$  can be done exactly, 
further analytic continuation to non-integer $n$ appears  to be ambiguous.
The classical example of this situation is provided by the Derrida's Random Energy Model
(REM) in which the momenta $Z(n)$ growths as $\exp(n^2)$ at large $n$, and in this case
there are many different distributions yielding the same values of $Z(n)$, but 
providing {\it different} values for the average free energy of the system \cite{REM}.
Performing "direct" analytic continuation to non-integer $n$
(just assuming that the parameter $n$ in the obtained expression for $Z(n)$ can take arbitrary
real values), one finds the so called replica symmetric (RS) solution which turns out to be 
correct at high temperatures, but which is apparently wrong (it provides negative entropy)
in the low temperature (spin-glass) phase. In the case of REM the situation is sufficiently
simple because here one can check what is right and what is wrong comparing with the 
available exact solution (which can be derived without replicas). Unfortunately in other systems 
the status of the results obtained by the replica method is much less clear. 

In the case of the mean-field spin-glasses \cite{SK} the replica partition function 
also growths as $\exp(n^2)$ at large $n$, and its "direct" analytic continuation
 to non-integer $n$,
as in REM, provides wrong RS solution in the low temperature spin-glass phase. Here
the solution which is generally {\it believed} to be correct is obtained via
the Parisi replica symmetry breaking (RSB) scheme 
(in the case of REM it reduces to the special case
which is called one-step RSB), and it is derived in terms of a {\it heuristic}
procedure  and not as a proper analytic continuation 
from integer to non integer values of $n$ of the replica partition function \cite{RSB-general}.
Recently the results obtained in terms of the RSB scheme has been confirmed by independent
mathematically rigorous calculations (see \cite{guerra} and references therein). 
Although no one seems to doubt now that the 
RSB heuristic procedure provide correct results, the problem is that until now no one was able to explain, 
{\it why} it provides correct results?

Presumably 
the most notable progress in the studies of the subtleties of the replica 
method has been achieved recently in the context of the random matrix 
theory, where the remarkable exact relation between replica partition functions
and Painlev\'e transcendents has been proved \cite{kanzieper1,splittorff,kanzieper2,kanzieper3}.
One can also mention here recent exact replica solution for one-dimensional directed polymers in random potential, which (unlike all previous examples) {\it did not} involve an analytic
continuation from integer to non-integer replica parameter $n$ and where the results obtained 
were also expressed in terms of the Painlev\'e transcendents \cite{Dotsenko,LeDoussal}.

In this paper I would like to consider two examples of the
systematic replica field theory calculations using very simple 
random systems for which, at first sight, every step of the replica program is under control.
It turns out, however, that even in these extremely simple cases the derivation 
of the physical results inevitably requires the usual replica method "cheating"
(the advantage of the simple system is that here one can easily see how 
it goes).  
Of course, it is not that present research explains something deep about the replica method.
The aim of the paper is (once again) to turn the attention 
to the existing paradoxes, and to promote (once again) the idea that "something has to be done":
it looks rather uncomfortable that on one hand we have extremely robust
method, which in most of the cases works perfectly well,
while on the other hand, we do not understand {\it why} it works.

\vspace{5mm}

The "toy" models considered in this paper are "extracted" from the one-dimensional 
directed polymers in a quenched random potential (for the detailed physical analysis of 
the obtained results see \cite{gaussian}). These systems  describe an elastic string  
directed along the $x$-axis within an interval $[0,L]$ with displacements  defined by 
the scalar field $\phi(x)$ having the elastic energy density proportional to $(\partial_x\phi)^2$. 
Randomness enters the problem through a disorder potential $V[\phi(x),x]$  competing against 
the elastic energy (see e.g. \cite{hh_zhang_95}).
The problem then is defined by the Hamiltonian
\begin{equation}
   \label{1gdp1}
   H[\phi(x),V] = \int_{0}^{L} dx
   \Bigl\{\frac{1}{2} \bigl[\partial_x \phi(x)\bigr]^2 
   + V[\phi(x),x]\Bigr\};
\end{equation}
The disorder potential $V[\phi(x),x]$ is Gaussian distributed
with a zero mean $\overline{V(\phi,x)}=0$ and a correlator
\begin{eqnarray}
   \label{1gdp2}
   {\overline{V(\phi,x)V(\phi',x')}} = \delta(x-x') U(\phi-\phi');
\end{eqnarray}
which is defined by a correlation function $U(\phi)$. The above equation implies that the 
random potential correlations are "translation invariant" in the $\phi$ direction 
depending only on the difference $(\phi - \phi')$.

General strategy of the replica calculations for this system 
is in the following. For the string with the zero boundary conditions
at  $x=0$  the partition function of a given sample is
\begin{equation}
\label{1gdp3}
   Z[V] = \int_{-\infty}^{+\infty} dy \int_{\phi(0)=0}^{\phi(L)=y} 
              {\cal D} [\phi(x)]  \;  \mbox{\Large e}^{-\beta H[\phi,V]}
\end{equation}
where the integration goes over all trajectories $\phi(x)$ staring at the 
origin, and $\beta$ denotes the inverse temperature. On the other hand,
the partition function is related to the total free energy $F[V]$ via
\begin{equation}
\label{1gdp4}
Z[V] = \exp( -\beta  F[V])
\end{equation}
The free energy $F[V]$  is defined for a specific 
realization of the random potential $V$ and thus represent a random variable. 
Let us take the $n$-th power of both sides of Eq.(\ref{1gdp4})
and perform the averaging over the random potential $V$:
\begin{equation}
\label{1gdp5}
Z[n,L] = \overline{\biggl(\exp( -\beta n F[V]) \Biggr)}
\end{equation}
The quantity in the l.h.s of the above equation
\begin{equation}
\label{1gdp6}
Z[n,L] \equiv \overline{\Bigl(Z[V]\Bigr)^{n}} 
\end{equation}
is called the replica partition function, and it is defined originally for an arbitrary {\it integer} parameter $n$. Let us suppose that at large $L$ 
the fluctuating free energy
of the system scales with the system size as $F \propto L^{\omega}$, 
i.e. it is characterized by a single universal exponent $\omega$. 
Redefining $F = f L^{\omega}$, we introduce a random quantity $f \sim 1$ which
can be  described by a distribution function ${\cal P}_{L}(f)$ (depending on the system size $L$).
In this way, instead of eq.(\ref{1gdp5}) we get  the following general relation 
between the replica partition function $Z[n,L]$ and the distribution function 
of the (rescaled) free energy fluctuations ${\cal P}_{L}(f)$:
\begin{equation}
\label{1gdp7}
   Z[n,L] \; =\;
           \int_{-\infty}^{+\infty} df \, {\cal P}_{L} (f) \;  
           \mbox{\Large e}^{ -\beta n  L^{\omega} \, f}
\end{equation}
The above equation is the bilateral Laplace transform of  the function ${\cal P}_{L}(f)$,
and at least formally it allows to restore this function in terms of the replica partition function $Z[n,L]$. In order to do so we have to compute $Z[n,L]$ for an {\it arbitrary} 
integer $n$ and then (if the result would permit!)
perform  analytical continuation of this function from integer 
to arbitrary complex values of $n$. 
Introducing  a new {\it complex} variable 
\begin{equation}
\label{1gdp8}
s = \beta n L^{\omega}
\end{equation}
and denoting
\begin{equation}
\label{1gdp9}
Z[\frac{s}{\beta L^{\omega}}, L] \equiv Z_{L}(s)
\end{equation}
instead of eq.(\ref{1gdp7}) we get
\begin{equation}
\label{1gdp7a}
   Z_{L}(s) \; =\;
           \int_{-\infty}^{+\infty} df \, {\cal P}_{L} (f) \;  
           \mbox{\Large e}^{ -s \, f}
\end{equation}
According to this relation the distribution function ${\cal P}_{L}(f)$ 
can be reconstructed via the inverse Laplace transform
\begin{equation}
\label{1gdp10}
   {\cal P}_{L}(f) \; = \;  \int_{-i\infty}^{+i\infty}  \frac{ds}{2\pi i}
       \; Z_{L}(s) \;   \mbox{\Large e}^{s f},
\end{equation}
where the integration goes over the contour parallel to the imaginary axis.
Finally, provided there are exist a {\it finite} thermodynamic limit function
\begin{equation}
\label{1gdp11}
\lim_{L\to\infty} Z_{L}(s) \; \equiv \; Z_{*}(s) 
\end{equation}
we can find the distribution function 
\begin{equation}
\label{1gdp12}
   {\cal P}_{*}(f) \; = \;  \int_{-i\infty}^{+i\infty}  \frac{ds}{2\pi i}
                \; Z_{*}(s) \;   \mbox{\Large e}^{s f},
\end{equation}
which would describe the statistics of the rescaled free energy fluctuations $f$ 
in the infinite system. The above equation defining ${\cal P}_{*}(f)$ contains no parameters 
and hence is expected to be universal. 
Note also that according to the relation $s = \beta n L^{\omega}$ 
in the thermodynamic limit $L\to\infty$ the relevant values of the  
replica parameter $n \sim L^{-\omega} \to 0$, which 
explains why the two limits $L\to\infty$ and $n\to 0$ do not commute \cite{kardar2}.

Usually the problem formulated in eqs.(\ref{1gdp1})-(\ref{1gdp2})
is studied in the context of the
short-range correlated disorder potential, 
i.e., for a rapidly decaying function $U(\phi \to \infty) \to 0$
(which for simplicity is often replaced by the $\delta$-function)
In this case  the free energy fluctuations  scale as $L^{1/3}$
\cite{kardar1,numer1,numer2,burgers,Dotsenko,LeDoussal}. 
On the other hand, if we would like to study  the statistics 
of {\it small} string displacements we could
develop the random potential $V[\phi,x]$ in powers of $\phi \ll 1$ 
keeping the first two terms only:
\begin{equation}
\label{1gdp13}
V[\phi,x] \simeq V_{0}(x) + V_{1}(x) \; \phi
\end{equation}
where $V_{0}(x)$ and  $V_{1}(x)$ are  the Gaussian uncorrelated random parameters with 
the zero mean, $\overline{V_{0}(x)} = \overline{V_{1}(x)} = 0$,
and the correlators
\begin{eqnarray}
\nonumber
\overline{V_{0}(x)V_{0}(x')} &=& v \; \delta(x-x') 
\\
\nonumber
\\
\label{1gdp14}
\overline{V_{1}(x)V_{1}(x')} &=& u \;     \delta(x-x') 
\\
\nonumber
\\
\nonumber
\overline{V_{1}(x)V_{0}(x')} &=& 0
\end{eqnarray}
which implies that unlike eq.(\ref{1gdp2}), the random potential correlator
is {\it not} translation invariant: 
\begin{eqnarray}
   \label{1gdp15}
   {\overline{V(\phi,x)V(\phi',x')}} = \delta(x-x')
\bigl[ v \; + \; u \phi \phi' \bigr]
\end{eqnarray}
In this way, instead of eq.(\ref{1gdp1}) we arrive to the Hamiltonian
\begin{equation}
\label{1gdp16}
H[\phi(x),V] = \int_{0}^{L} dx
   \Bigl\{\frac{1}{2} \bigl[\partial_x \phi(x)\bigr]^2 
   + V_{1}(x) \phi(x) + V_{0}(x)\Bigr\};
\end{equation}
Now,  lifting the requirement $\phi \ll 1$,  we are getting simple Gaussian random force model 
to be studied in this paper by the replica method (Section II). 

The thermodynamic limit,
$L\to\infty$, of this system has been studied earlier \cite{gorokhov-blatter}.
Here I would like to concentrate on the thechnical details of the calculations.
Due to apperent simplicity of the model the replica trick calculations
can be followed here step by step from the very begining till the very end. 
In this way it can be easily demonstrated that {\it formally} at sertain stage of the
calculations the implementation of the replica program (as it is declared at the very begining)
is just impossible. On the other hand, following the 
usual "doublethink"  traditions  of the replica calculations in disordered systems 
(i.e. closing eyes on the fact that certain suggestions used in the calculations
contradict to each other) one can easily fulfill the programme till the very end
to obtain very nice and physically sensible result for the free energy distribution function.
Moreover, in this particular case we can be sure that the result obtained in this way
is indeed {\it correct}, as it can also be derived via direct calculations
without replicas \cite{gaussian}.
It turns out that regardless of the apparent simplicity of the model, its 
free energy distribution function (which will be derived here for an arbitrary finite 
system size $L$) is rather non-trivial. 

\vspace{5mm}

In Section III we consider slightly modified version of the above "toy" system.
Namely, instead of developing the random potential itself, eq.(\ref{1gdp13}),
one can consider the development of its {\it correlation function} $U(\phi)$,
eq.(\ref{1gdp2}). Again, keeping the first two terms only, one gets
\begin{equation}
   \label{1gdp17}
U(\phi) \; \simeq \;  v - \frac{1}{2} u \phi^{2}
\end{equation}
(which, unlike eq.(\ref{1gdp15}), would preserve the translation invariance 
of the random potential correlations).
It turns out that this, seemingly rather innocent modification 
 provokes quite dramatic consequences. As before, we 
lift the requirement $\phi \ll 1$ and accept the above "truncated"
correlation function, eq.(\ref{1gdp17}), as valid for the whole range
of the scalar fields $\phi(x)$.  Then, in the result of the 
standard replica calculations  (which again inevitably involve the same 
cheeting as in the case of the previous model)
we find that the corresponding free energy distribution function  
is {\it not positively defined}, which, of course,  makes no physical sense.
The point of this little methodological (and pedagogical) exercise is to demonstrate
that it is not the replicas which are always responsible for all the troubles
in the disordered systems world.
In fact, the result of the replica calculations honestly reproduces the
pathological nature of the original model itself: due to approximation,
eq.(\ref{1gdp17}), we obtain the replica theory which does not correspond
to any physical system. One can  show that
there exists no (positively defined) Gaussian distribution function of the 
random potentials $V[\phi,x]$ which would provide the correlation function,
eqs.(\ref{1gdp2}), in the "parabolic" form, eq.(\ref{1gdp17}). 
Moreover, the proper definition of the disorder potential is subject to important
general constraints \cite{denis_99} regarding the shape of the correlation function $U(\phi)$, 
and neglection of these constraints may lead to unphysical results.

Indeed, consider a random potential $V(\phi)$ and its Fourier 
representation $\tilde{V}(p) = \int d\phi \; V(\phi) \exp(-ip\phi)$.
Then the Gaussian distribution function of the random function
$\tilde{V}(p)$ has the form 
\begin{equation}
   \label{1gdp18}
   {\cal P}[\tilde{V} (p)] = P_0 \exp\Bigl(-\int \frac{dp}{2\pi}\,
   \frac{|\tilde{V}(p)|^2}{2 G(p)}\Bigr);
\end{equation}
where the {\it positive} function  $G(p)$ is related with the correlation function 
$U(\phi)$ via 
\begin{equation}
   \label{1gdp19}
   U(\phi) = \int \frac{dp}{2\pi} \, G(p) \,\exp(i p\phi).
\end{equation}
Expanding both sides of the above relation in powers of $\phi$, 
\begin{equation}
   \label{1gdp20}
   U(0) + \sum_{k=1}^{\infty} \frac{1}{(2k)!}U^{(2k)}(0)\phi^{2k} =
   \int \frac{dp}{2\pi} \, G(p) 
   + \sum_{k=1}^{\infty} \frac{(-1)^{k}}{(2k)!}
   \Bigl(\int \frac{dp}{2\pi} G(p)\, p^{2k}\Bigr)\phi^{2k},
\end{equation}
we can compare coefficients: the $2k$-th derivative of $U(\phi)$
in the origin relates to the integral $\int dp G(p) p^{2k}$
which is a positive quantity. Hence, we have to be careful in our 
choice of the correlator $U(\phi)$: if we truncate the expansion
of $U(\phi)$ beyond some $k^*$, such that $U^{(2k)}(0) = 0$ for $k 
\geq k^*$, we impose the condition 
\begin{equation}
   \label{1gdp21}
   \int dp G(p) p^{2k} = 0 \; \; \textrm{  for  } k\geq k^*,
\end{equation}
which cannot be satisfied for a positively defined $G(p)$. 
Obviously, choosing the correlator $U(\phi)$, in the "parabolic" form,
eq.(\ref{1gdp17}) is in severe conflict with this constraint. 
If nevertheless, we would substitute such a parabolic correlator
into the replica Hamiltonian and perform all the standard
calculations, first we would see no apparent indications 
telling that something went wrong.
Moreover, one can easily calculate the average free energy of this system 
to discover that it perfectly coincides with the one of the (physically consistent) 
random force model, eq.(\ref{1gdp16}).
The trouble appears in the calculation of more specific quantities.
For example, the second cumulant of free energy fluctuations 
turns out to be negative  which makes no physical sense. 
Including the next term $\propto \phi^4$ in the
correlator's expansion can cure this problem, however, an
inconsistency then shows up in the next order cumulants, etc.

\vspace{5mm}

The paper is organized as follows. In  Section II
 we perform detailed analycis of the systematic replica field theory calculations for the random force
model, eq.(\ref{1gdp16}), and present the results for its free energy distribution function.
In Section III similar solution is considered for the "parabolic" directed polymer problem
described by the correlation function, eq.(\ref{1gdp17}), providing
not positively defined free energy distribution function.
Finally,  the subtleties of the replica method
are discussed in Section IV.

\section{Random force model}

Explicitly, the replica partition function, Eq.(\ref{1gdp6}), of the system described by
the Hamiltonian, Eq.(\ref{1gdp16}), is

\begin{equation}
\label{2gdp1}
   Z(n,L) = \prod_{a=1}^{n} \int_{-\infty}^{+\infty} dy_{a} 
                            \int_{\phi_{a}(0)=0}^{\phi_{a}(L)=y_{a}} 
   {\cal D} \phi_{a}(x) \;
   \overline{\exp\Biggl[-\beta \int_{0}^{L} dx \sum_{a=1}^{n}
   \bigl\{\frac{1}{2} \bigl[\partial_x \phi_{a}(x)\bigr]^2 
   + V_{1}(x) \phi_{a}(x) + V_{0}(x)   \bigr\}\Biggr] }
\end{equation}
Since  the random parameters $V_{1}(x)$ and $V_{0}(x)$ have Gaussian 
distribution with correlations defined in eq.(\ref{1gdp14}), the disorder average $\overline{(...)}$ in the above equation is very simple:

\begin{equation}
\label{2gdp2}
 \overline{\exp\Biggl[-\beta \int_{0}^{L} dx \sum_{a=1}^{n} \bigl\{
 V_{1}(x) \phi_{a}(x) + V_{0}(x)   \bigr\}\Biggr] } \; = \; 
\exp\Biggl[\frac{1}{2} \beta^{2} \int_{0}^{L} dx \sum_{a,b=1}^{n} \bigl\{
u \phi_{a}(x) \phi_{b}(x) + v \bigr\}\Biggr]
\end{equation}
Thus, the replica partition function, Eq.(\ref{2gdp1}), 
can be represented in the following form
\begin{equation}
   \label{2gdp3}
   Z(n,L) \; = \;  \mbox{\Large e}^{ \frac{1}{2} \beta^{2} n^{2} v L }
\prod_{a=1}^{n} \int_{-\infty}^{+\infty} dy_{a} \; 
\Psi\bigl[{\bf y}; L\bigr]
\end{equation}
where the "wave function"
\begin{equation}
   \label{2gdp4}
\Psi\bigl[{\bf y}; L\bigr] \; = \; 
\int_{\phi_{a}(0)=0}^{\phi_{a}(L)=y_{a}} 
   {\cal D} \phi_{a}(x) \;
   \exp\Bigl[-\beta H_{n}[{\boldsymbol \phi}]  \Bigr]
\end{equation}
is defined by  $n$-component scalar fields replica Hamiltonian
\begin{eqnarray}
\nonumber
   H_{n}[{\boldsymbol \phi}] &=&  
   \frac{1}{2} \int_{0}^{L} dx \Biggl(
   \sum_{a=1}^{n} \bigl[\partial_x \phi_{a}(x)\bigr]^2 
   - \beta u \sum_{a, b}^{n} \phi_{a}(x) \phi_{b}(x) \Biggr) 
\\
\nonumber
\\
&=& 
-\frac{1}{2} \int_{0}^{L} dx \sum_{a,b=1}^{n} \phi_{a}(x) \, U_{ab} \, \phi_{b}(x)
\label{2gdp5}
\end{eqnarray}
The matrix
\begin{equation}
\label{2gdp6}
U_{ab} =   \; \partial^{2}_x \, \delta_{ab} \; + \; \beta u 
\end{equation}
can be easily diagonalized. It has $(n-1)$-degenerate eigenvalue

\begin{equation}
\label{2gdp7}
\lambda_{1} \; = \;   \partial^{2}_x
\end{equation}
with $(n-1)$ orthonormal eigenvectors $\xi^{a}_{i}$ such that
\begin{equation}
\label{2gdp8}
\sum_{a=1}^{n} \xi^{a}_{i} = 0 \; \; \; \; \; (i = 1, ..., n-1)
\end{equation}
and one non-degenerate eigenvalue
\begin{equation}
\label{2gdp9}
\lambda_{2} \; = \; \partial^{2}_x \, + \, \beta n u
\end{equation}
with the eigenvector 
\begin{equation}
\label{2gdp10}
\xi^{a}_{n} = 1/\sqrt{n} 
\end{equation}
Thus, the matrix $U_{ab}$, Eq.(\ref{2gdp6}), is diagonalized by the 
orthonormal transformation which is defined by the $(n\times n)$ matrix 
$\xi^{a}_{i}$ (such that $\sum_{a=1}^{n} \xi^{a}_{i} \xi^{a}_{j} = \delta_{ij}$ and
$\sum_{i=1}^{n} \xi^{a}_{i} \xi^{b}_{i} = \delta_{ab}$).
In terms of new fields
\begin{equation}
\label{2gdp11}
\varphi_{i}(x) \; = \; \sum_{a=1}^{n} \, \xi^{a}_{i} \, \phi_{a}
\end{equation}
the Hamiltonian, Eq.(\ref{2gdp5}), takes the form
\begin{equation}
   \label{2gdp13}
H_{n}[{\boldsymbol \varphi}] \; = \; 
\frac{1}{2} \int_{0}^{L} dx \sum_{i=1}^{n-1}\bigl[\partial_x \varphi_{i}(x)\bigr]^2 + 
\frac{1}{2} \int_{0}^{L} dx \Bigl\{ \bigl[\partial_x \varphi_{n}(x)\bigr]^2 
  - \beta n u \varphi_{n}^{2}(x) \Bigr\}
\end{equation}
with the boundary conditions
\begin{equation}
\label{2gdp12}
\varphi_{i}(L) \; \equiv \; r_{i}({\bf y}) \; = \; \sum_{a=1}^{n} \, \xi^{a}_{i} \, y_{a}
\end{equation}
Correspondingly, the wave function, eq.(\ref{2gdp4}), factorizes into 
\begin{equation}
\label{2gdp14}
\Psi\bigl[{\bf y}; L\bigr] \; = \; 
\Biggl[\prod_{i=1}^{n-1} \Psi_{0}\bigl[r_{i}({\bf y}); L\bigr] \Biggr] \; 
\Psi_{1}\bigl[r_{n}({\bf y}); L\bigr]
\end{equation}
where (with the proper choice of the integration measure)
\begin{equation}
\label{2gdp15}
 \Psi_{0} (r; L) \; = \;  \int_{\varphi(0)=0}^{\varphi(L)=r} {\cal D} \varphi(x) \; 
             \exp\biggl(-\frac{1}{2} \beta  \int_{0}^{L} dx
             \bigl[\partial_x \varphi(x)\bigr]^2 \biggr) \; = \; 
             \sqrt{\frac{\beta }{2\pi L}} \; \exp\biggl(-\frac{\beta }{2 L} r^{2} \biggr)
\end{equation}
 and 
\begin{equation}
\label{2gdp16}
\Psi_{1}(r; L) \; = \; \int_{\varphi(0)=0}^{\varphi(L)=r} {\cal D} \varphi(x) \; 
                  \exp\biggl(-\frac{1}{2} \int_{0}^{L} dx 
                  \bigl[ \beta  \bigl(\partial_x \varphi(x)\bigr)^2 
                       - \beta^{2} n u \varphi^{2}(x)\bigr] \biggr)
\end{equation}
According to the above definition the wave function $\Psi_{1}(r,t)$ satisfies 
the imaginary-time Schr\"odinger equation
\begin{equation}
   \label{2gdp17}
   \partial_t \Psi(r; t) \; = \; \frac{1}{2\beta} \partial_{r}^2 \Psi_{1}(r; t)\;
                              + \; \frac{1}{2} \, \beta^{2} n u  \, r^{2} \, \Psi_{1}(r; t)
\end{equation}
with the initial condition
\begin{equation}
   \label{2gdp18}
   \Psi_{1}(r; t=0) \; = \; \delta(r)
\end{equation}
Eq.(\ref{2gdp17}) describes the movement of a particle in the "reversed" parabolic potential.
One can easily check that the solution of this equation satisfying the above initial condition is
\begin{equation}
   \label{2gdp19}
\Psi_{1}(r; t) \; = \; b(t)  \, \exp\biggl(-\frac{1}{2} a(t) \, r^{2} \biggr)
\end{equation}
where
\begin{equation}
   \label{2gdp20}
b(t) \; = \; \sqrt{\frac{\beta }{2\pi t}} \, 
                \frac{(\lambda t^{2})^{1/4}}{\sqrt{\sin(\sqrt{\lambda t^{2}})}}
\end{equation}
and
\begin{equation}
   \label{2gdp21}
a(t) \; = \; \beta  \, \sqrt{\lambda} \, 
                \frac{\cos(\sqrt{\lambda t^{2}})}{\sin(\sqrt{\lambda t^{2}})}
\end{equation}
where we have introduced the parameter
\begin{equation}
   \label{2gdp22}
\lambda \; = \; \beta n u
\end{equation}
It should be stressed that the above solution exists  provided
\begin{equation}
   \label{2gdp23}
0 \; < \; \lambda t^{2} \; < \; \frac{\pi^{2}}{4}
\end{equation}
This restriction indicates that at a given value of the parameter $\lambda$ the elastic string 
described by the partition function, eq.(\ref{2gdp16}) (which contains negative mass!) goes
to infinity at the {\it finite} time $t_{c} = \pi/(2 \sqrt{\lambda} )$. Coming back to the 
 original random force problem  one can reformulate
the above restriction in the other way: for a given system size ("time") $L$ 
the corresponding replica partition function $\overline{Z^{n}}$  is defined 
only if 
\begin{equation}
   \label{2gdp24}
n \; < \; n_{c}(L) \; = \; \frac{\pi^2}{4 \beta u L^{2}}
\end{equation}
At bigger values of $n$ the replica partition function is simply  not defined 
(it is formally divergent). Taking into account that the quantity
$\pi^2/\bigl(4 \beta u L^{2}\bigr)$ can be easily made less than one (e.g. by taking
$L$ sufficiently large),
while the replica parameter $n$ is still should be kept positive integer, we see that 
the above restriction becomes the fatal point for the whole replica scheme of the 
calculations. Of course, it is tempting to claim that since the replica $n$ enter
the partition function, eq.(\ref{2gdp16}), in the form of the analytic parameter
$\beta^{2} n u$, we can analytically continue it to arbitrary real values in
the interval $0 < n < n_{c}$, eq.(\ref{2gdp24}). But the problem is that the expression for the 
whole replica partition function, eqs.(\ref{2gdp3}), (\ref{2gdp14}), requires
that the replica parameter $n$ {\it must} be still integer. 
In fact, the restriction, eq.(\ref{2gdp24}), reflects simple mathematical reality:
the replica partition function $\overline{Z^{n}}$ of the considered system
is divergent at $n \geq n_{c}(L)$. In particular, at $L \geq \pi/\sqrt{4\beta u}$,
the replica partition function does no exist (divergent) for {\it all} integers $n =1, 2, ...$.
In this situation (from the point of view of the common sense) the application 
of the replica method program for the system under consideration 
looks completely impossible. Nevertheless, in terms of the  {\it modus vivendi}  of the
replica method this problem is overcome in a very simple way. It is in this point that
the "doublethink" begins: wherever the replica $n$ appears in a form of an analytic parameter
it is immediately considered as real and belonging to the desired region ($n\to 0$ in spin glasses,
or $0 < n < n_{c}$ in the present case), while {\it at the same time} wherever $n$ 
can not made non-integer (in the summations or in the products) it is still considered
as an integer (note that similar, although slightly more sophisticated trick is used in the 
replica symmetry breaking construction). Thus,  we continue our
calculations just plainly assuming that the condition, eq.(\ref{2gdp24}), is satisfied.

Substituting, the solutions eq.(\ref{2gdp15}), (\ref{2gdp19})-(\ref{2gdp21}) into 
eq.(\ref{2gdp14}) ,  we get
\begin{equation}
\label{2gdp25}
\Psi\bigl[{\bf y}; L\bigr] \; = \; 
\sqrt{\frac{\sqrt{\lambda L^{2}}}{\sin(\sqrt{\lambda L^{2}})}}
\Biggl(\frac{\beta}{2\pi L}\Biggr)^{n/2 }
\exp\Biggl[ -\frac{\beta}{2 L} \sum_{i=1}^{n-1} r_{i}^{2}\bigl({\bf y}\bigr)
-\frac{\beta}{2L} 
 \frac{\sqrt{\lambda L^{2}} \cos(\sqrt{\lambda L^{2}})}{\sin(\sqrt{\lambda L^{2}})} 
r_{n}^{2}\bigl({\bf y}\bigr) \Biggr]
\end{equation}
Using the relations,
eqs.(\ref{2gdp12}), (\ref{2gdp8}) and (\ref{2gdp10}), and taking into account that the matrix $\xi_{i}^{a}$ 
is orthonormal,  after some efforts in simple algebra
we obtain the following result for the $n$-particle wave function, eq.(\ref{2gdp4}):
\begin{equation}
\label{2gdp26}
\Psi\bigl[{\bf y}; L\bigr] \; = \; 
\sqrt{\frac{\sqrt{\lambda L^{2}}}{\sin(\sqrt{\lambda L^{2}})}}
\Biggl(\frac{\beta}{2\pi L}\Biggr)^{n/2 }
\exp\Biggl[ -\frac{\beta}{2 L} \sum_{a=1}^{n} y_{a}^{2} \; - \; 
\frac{\beta}{2 L n} \Biggl(
\frac{\sqrt{\lambda L^{2}} \cos(\sqrt{\lambda L^{2}})}{\sin(\sqrt{\lambda L^{2}})} 
- 1 \Biggr) \Bigl(\sum_{a=1}^{n} y_{a} \Bigr)^{2} \Biggr]
\end{equation}
where it assumed that ({\it sic}!) $\lambda L{^2} = \beta n u L^{2} < \pi^{2}/4$ whatever 
the values of $\beta, u$ and  $L$ are.
Substituting this result into eq.(\ref{2gdp3}) and performing simple Gaussian integration
(which, taking into account orthogonality of the matrix $\xi_{i}^{a}$ is easier
to do in terms of the parameters $r_{i}$ using expression, eq.(\ref{2gdp25})),
for the replica partition function we finally get sufficiently simple result
\begin{equation}
   \label{2gdp27}
Z(n,L) \; = \;  \frac{1}{\sqrt{\cos(\sqrt{\beta n u  L^{2}})}} \; 
\exp\Bigl[ \frac{1}{2} \beta^{2} n^{2} v L \Bigr]
\end{equation}
Next, for further implementation of the general program of the 
reconstruction of the free energy distribution function as it was described in the 
Introduction, eqs.(\ref{1gdp3})-(\ref{1gdp10}), 
let us introduce parameter
\begin{equation}
   \label{2gdp28}
w \; = \; \beta n u L^{2} 
\end{equation}
which is confined in the interval $0 < w < \pi^{2}/4$. In terms of this parameter
the general relation between the replica partition function 
$Z(n,L)$ and the free energy distribution function $P_{L}(F)$ (cf. eq.(\ref{1gdp5})),
\begin{equation}
   \label{2gdp29}
Z(n,L) \; = \;  \int_{-\infty}^{+\infty} dF \;  P_{L}(F) \; \exp\bigl(-\beta n F\bigr)
\end{equation}
takes the form
\begin{equation}
   \label{2gdp30}
Z_{L}(w) \; = \;  \int_{-\infty}^{+\infty} dF \;  P_{L}(F) \; \exp\bigl(-\frac{w}{u L^{2}}
 F\bigr)
\end{equation}
where
\begin{equation}
   \label{2gdp31}
Z_{L}(w) \equiv Z\bigl(\frac{w}{\beta u L^{2}},L) 
\; = \;  \frac{1}{\sqrt{\cos(\sqrt{w})}} \; 
\exp\Bigl[ \frac{v}{2u^{2} L^{3}}w^{2} \Bigr]
\end{equation}
Rescaling the free energy, $F \; = \; u L^{2} f$, for the probability distribution function 
${\cal P}_{L}(f)$ of the random quantity $f$ we get the following relation
\begin{equation}
   \label{2gdp32}
Z_{L}(w) \; = \;  \int_{-\infty}^{+\infty} df \;  {\cal P}_{L}(f) \; 
\exp\bigl(- w  f \bigr)
\end{equation}
(where ${\cal P}_{L}(f) \; = \; uL^{2} P_{L}\bigl(uL^{2} f\bigr)$). 
Next, performing the analytic continuation of the function $Z_{L}(w)$, eq.(\ref{2gdp31}),
from the interval $0 < w < \pi^{2}/4$ to the complex half-plain, Re$\{w\} < \pi^{2}/4$,
(which is unambiguous operation) the free energy distribution function ${\cal P}_{L}(f)$
can be obtained via the inverse Laplace transform
\begin{equation}
\label{2gdp33}
   {\cal P}_{L}(f) =  \int_{-i\infty}^{+i\infty} \frac{dw}{2\pi i}
  \, Z_{L}(w) \, \exp( w f),
\end{equation}
where the integration goes over the contour parallel to the imaginary axes
such that Re$\{w\} < \pi^{2}/4$.

In the thermodynamic limit,
according to eq.(\ref{2gdp31}), 
\begin{equation}
   \label{2gdp34}
\lim_{L\to\infty} Z_{L}(w) \equiv Z_{*}(w)  
\; = \;  \frac{1}{\sqrt{\cos(\sqrt{w})}} \; 
\end{equation}
Thus, according to eq.(\ref{2gdp33}),
for the  distribution function of the rescaled free energy fluctuations of the infinite system,
${\cal P}_{*}(f) \equiv \lim_{L\to\infty} {\cal P}_{L}(f)$,
we obtain the following (universal) result \cite{gorokhov-blatter}:
\begin{equation}
\label{2gdp35}
   {\cal P}_{*}(f) =  \int_{-i\infty}^{+i\infty} \frac{dw}{2\pi i}
  \,\frac{\exp( w f)}{\sqrt{\cos(\sqrt{w})}}
\end{equation}

\begin{figure}[h]
\begin{center}
   \includegraphics[width=12.0cm]{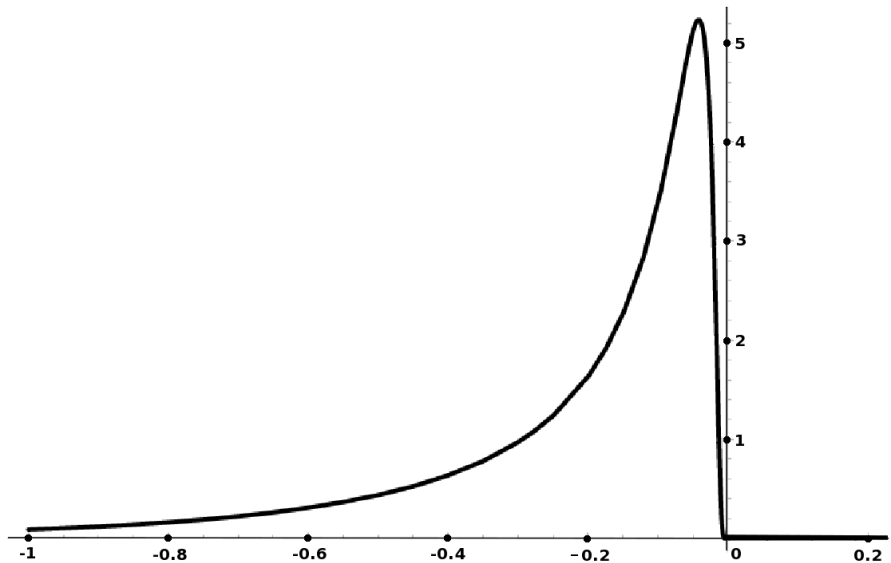}
\caption[]{The thermodynamic limit free energy distribution function ${\cal P}_{*}(f)$}
\end{center}
\label{figure1}
\end{figure}

The overall form of this function is shown in Figure 1. It is interesting to note
that this function is identically equal to zero at $f > 0$. This is easy to understand
by making simple mathematical analysis of the integral in eq.(\ref{2gdp35}). 
Indeed, since at $f>0$ the function $[\cos(\sqrt{w})]^{-1/2} \exp(wf)$ quickly
goes to zero at $w \to -\infty$, the contour of integration in the complex plane 
can be safely shifted to $-\infty$, which means that $ {\cal P}_{*}(f < 0) \equiv 0$.
The fact that in the thermodynamic limit the upper bound for the free energy 
of the system described by the Hamiltonian, eq.(\ref{1gdp16})
is equal to zero can also be explained in terms of simple physical arguments.
First we note that the typical value of the random constant term
$\int dx V_{0}(x)$ scales as $L^{1/2}$, which means that its contribution to
the rescaled free energy $f \sim F/L^{2}$ scales as $L^{-3/2}$ and vanishes 
in the thermodynamic limit $L\to\infty$. On the hand, since the 
contribution of the "trivial" configuration $\phi(x) = 0$ in the elastic 
and the random force terms of the Hamiltonian is equal to zero, 
any deviation  from this configuration (due to the actions of the random force) 
can only reduce the energy. The asymptotic behavior of the function ${\cal P}_{*}(f)$ in the limits
$f\to -\infty$ and $f\to -0$ can be easily estimated by the saddle-point integration
to yield: ${\cal P}_{*}(f\to -\infty)  \sim  \exp\bigl(-\frac{\pi^{2}}{4} |f|\bigr)$
and ${\cal P}_{*}(f\to -0)  \sim  \exp\bigl(-\frac{1}{32 |f|} \bigr)$.

At finite system size $L$, according to eqs.(\ref{2gdp31}) and (\ref{2gdp33}),
the free energy distribution function is given by
\begin{equation}
\label{2gdp39}
   {\cal P}_{L}(f) =  \int_{-i\infty}^{+i\infty} \frac{dw}{2\pi i}
  \,\frac{1}{\sqrt{\cos(\sqrt{w})}} \; \exp\Bigl[ \epsilon(L) w^{2} + f w \Bigr]
\end{equation}
where
\begin{equation}
\label{2gdp40}
\epsilon(L) \; = \; \frac{v}{2u^{2} L^{3}}
\end{equation}
The plot of this function  for several values of the parameter $\epsilon(L)$
is shown in Figure 2. At small values of $\epsilon$ (large $L$) the function 
$ {\cal P}_{L}(f)$ becomes close to the universal distribution function
$ {\cal P}_{*}(f)$ shown in Figure 1, while at large $\epsilon$ (when  $v \gg u^{2} L^{3}$)
$ {\cal P}_{L}(f)$ becomes almost Gaussian, as in this case 
the  free energy of the system is dominated by the Gaussian random constant term
$\int dx V_{0}(x)$ of the Hamiltonian,  eq.(\ref{1gdp16}).

\begin{figure}[h]
\begin{center}
   \includegraphics[width=12.0cm]{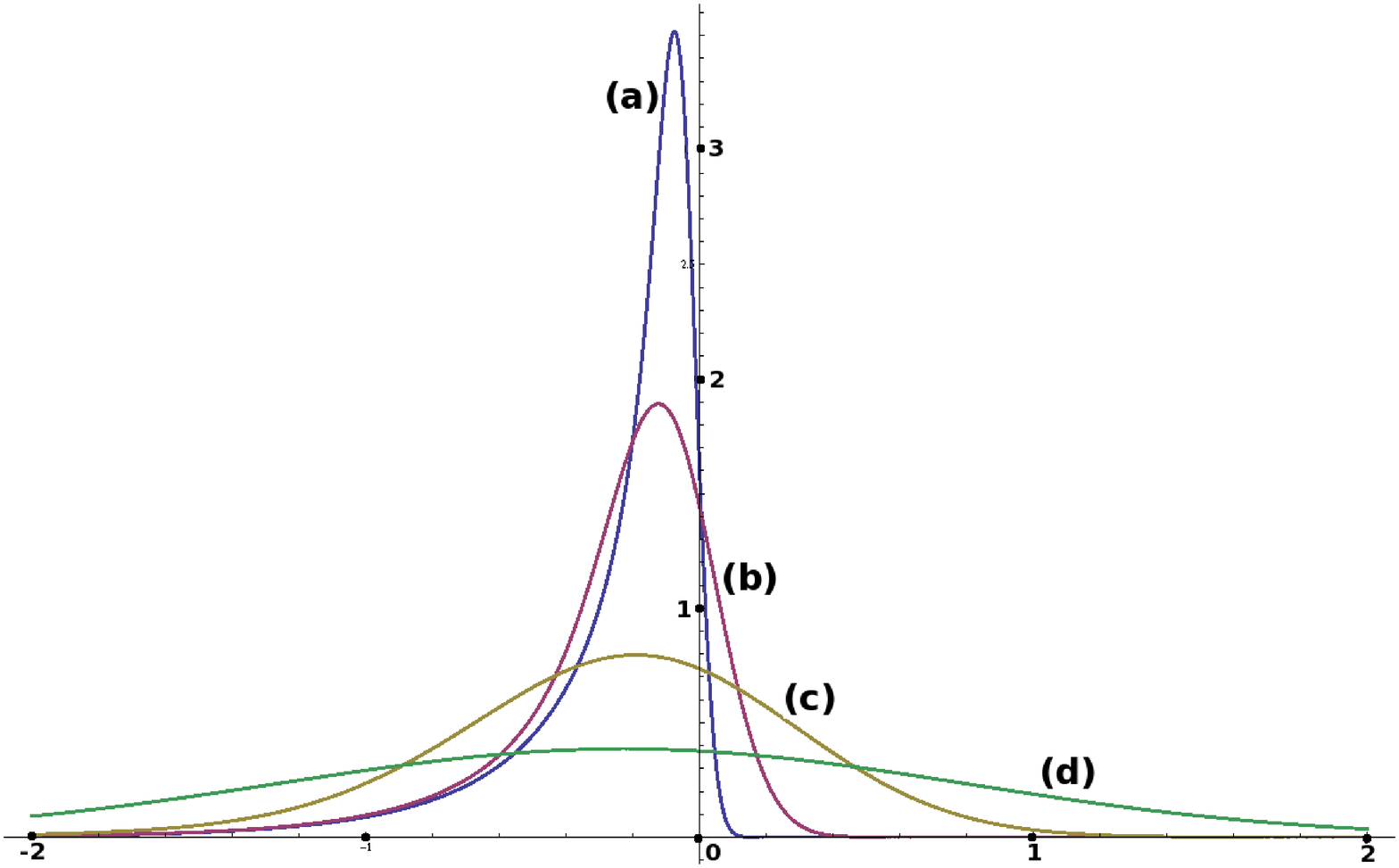}
\caption[]{ Free energy distribution function ${\cal P}_{L}(f)$, eq.(\ref{2gdp39}),
for: (a) $\epsilon =0.001 $, (b) $\epsilon =0.01 $, (c) $\epsilon =0.1 $, (d) $\epsilon =0.5 $,}
\end{center}
\label{figure2}
\end{figure}

The result, eq.(\ref{2gdp39}), constitutes the complete solution of the random force problem
defined by the Hamiltonian, eq.(\ref{1gdp16}). Could we call this solution "exact"?
One one hand, the explicit cheating in the the derivation of the replica partition function, eq.(\ref{2gdp27}),
makes the status of the obtained result rather indefinite. On the other hand,
it should be stressed that this kind free handling with the integer/non-integer status of the 
replica parameter $n$ is just the routine trick in all replica field theory calculations
in disordered systems (see e.g. \cite{RSB-general,replicas}). In other words, the results abtained in this way
should be accompanied by the label "in the framework of the replica approach". 
Fortunately, in this particular case, due to Gaussian nature of the considered model the result
for the replica partition function, eq.(\ref{2gdp27}), (where $n \in \bigl(0, n_{c}(L)\big)$
is a {\it real} parameter) can be confirmed by independent calculation without the use of replicas
\cite{gaussian}, which allows to claim that the above result, eq.(\ref{2gdp39}), for the free energy
distribution function is indeed exact.

\section{Directed polymers with "parabolic" correlations of the random potential}

The replica partition function of directed polymers described by the Hamiltonian (\ref{1gdp1})
with non-local parabolic correlations of the random potentials, Eq.(\ref{1gdp17}),
can also be represented in the form of eqs.(\ref{2gdp3})-(\ref{2gdp4}) with the 
replica Hamiltonian
\begin{eqnarray}
\nonumber
   \tilde{H}_{n}[{\boldsymbol \phi}] &=&  
   \frac{1}{2} \int_{0}^{L} dx \Biggl(
   \sum_{a=1}^{n} \bigl[\partial_x \phi_{a}(x)\bigr]^2 
   +\frac{1}{2} \beta u \sum_{a, b}^{n} \Bigl[\phi_{a}(x) - \phi_{b}(x) \Bigr]^{2} \Biggr) 
\\
\nonumber
\\
&=& 
-\frac{1}{2} \int_{0}^{L} dx \sum_{a,b=1}^{n} \phi_{a}(x) \, \tilde{U}_{ab} \, \phi_{b}(x)
\label{3gdp1}
\end{eqnarray}
with the matrix
\begin{equation}
\label{3gdp2}
\tilde{U}_{ab} =   \; \Bigl(\partial^{2}_x \, - \, \beta n u\Bigr) \delta_{ab} \; + \; \beta u 
\end{equation}
Following the same route as in the case of the random force model, eqs.(\ref{2gdp6})-(\ref{2gdp13}),
we note that the above matrix has $(n-1)$-degenerate eigenvalue $\tilde{\lambda}_{1} = \partial^{2}_x - \beta n u$
with $(n-1)$ orthonormal eigenvectors $\xi^{a}_{i}$ constrained by the condition
$\sum_{a=1}^{n} \xi^{a}_{i} =0 \; (i = 1, ..., n-1)$, 
and one non-degenerate eigenvalue $\tilde{\lambda}_{2} = \partial^{2}_x $ with the eigenvector 
$\xi^{a}_{n} = 1/\sqrt{n} $. 
In terms of the new fields $\varphi_{i}(x) = \sum_{a=1}^{n} \xi^{a}_{i}  \phi_{a}$,
the replica Hamiltonian take the form (cf. eq.(\ref{2gdp13}))
\begin{equation}
   \label{3gdp3}
\tilde{H}_{n}[{\boldsymbol \varphi}] \; = \; 
\frac{1}{2} \int_{0}^{L} dx \sum_{i=1}^{n-1}
\Bigl\{  \bigl[\partial_x \varphi_{i}(x)\bigr]^2 - \beta n u \phi_{i}^{2}(x) \Bigr\}
+ \frac{1}{2} \int_{0}^{L} dx \bigl[\partial_x \phi_{n}(x)\bigr]^2 
\end{equation}
Similarly  to the calculations of the previous section, eqs(\ref{2gdp14})-(\ref{2gdp27}),
for the replica partition function we get the following result (cf. eq.(\ref{2gdp27})):
\begin{equation}
   \label{3gdp4}
\tilde{Z}(n,L) \; = \;  \Biggl[\frac{1}{\sqrt{\cos(\sqrt{\beta n u  L^{2}})}}\Biggr]^{(n-1)}\; 
\exp\Bigl[ \frac{1}{2} \beta^{2} n^{2} v L \Bigr]
\end{equation}
where, as in the case of the random force model, it is assumed that $\beta n u L^{2} < \pi^{2}/4$. 
In terms of the parameter $w = \beta n u L^{2}$ instead of eq.(\ref{2gdp31}) we obtain
\begin{equation}
\label{3gdp5}
\tilde{Z}_{L,\beta}(w)  = \sqrt{\cos(\sqrt{w})} \; \biggl(\sqrt{\cos(\sqrt{s})}\biggr)^{-\frac{w}{2\beta u L^{2}}}
\exp\Bigl[ \frac{v}{2u^{2} L^{3}} w^{2} \Bigr]
\end{equation} 
and the probability distribution function of the (rescaled) free energy 
fluctuation $\tilde{{\cal P}}_{L}(f)$ is given by the inverse Laplace transform (cf. eq.(\ref{2gdp33})):
\begin{equation}
\label{3gdp6}
   \tilde{{\cal P}}_{L,\beta}(f) =  \int_{-i\infty}^{+i\infty} \frac{dw}{2\pi i}
  \, \tilde{Z}_{L}(w) \, \exp( w f),
\end{equation}
Performing (numerical) integration in the above equation one can easily find that
this function is {\it not positively defined} for any values of the parameters 
$\epsilon = v/(2u^{2} L^{3})$ and $\kappa = 1/(2\beta u L^{2})$. For example, 
in the zero temperature limit 
(when $\kappa \to 0$) the above equation reduces to
(cf. eq.(\ref{2gdp39})):
\begin{equation}
\label{3gdp7}
   \lim_{\beta\to\infty}\tilde{{\cal P}}_{L,\beta}(f) \; \equiv \; \tilde{{\cal P}}^{*}_{L}(f) 
\; = \; \int_{-i\infty}^{+i\infty} \frac{dw}{2\pi i}
  \,\sqrt{\cos(\sqrt{w})} \; \exp\Bigl[ \epsilon(L) w^{2} + f w \Bigr]
\end{equation}
The plot of this function  for several values of the parameter $\epsilon(L)$
is shown in Figure 3. We see that unlike  ${\cal P}_{L}(f)$ (Figure 2) 
of the random force model, the function $\tilde{{\cal P}}^{*}_{L}(f)$ even 
at small $L$ (large $\epsilon$) (when it is almost Gaussian) has always a kind of the negalive "kink" 
at sufficiently large $f$. 

Unfortunately we are not able to confirm (or reject)
the result, eq.(\ref{3gdp4}), by independent calculations, as the replica theory
defined by the Hamiltonian, eq.(\ref{3gdp1}), does not corresponds to any physical 
system (see the discussion of this issue in the Introduction, eqs.(\ref{1gdp18})-(\ref{1gdp21})).
On the other hand, the clear lesson which we can learn from the exercise considered in this Section, 
is that unlike the "honest cheating" with the status of the replica parameter $n$
(discussed in previous Section), any {\it approximations} made at the stage of the replica calculations
(the exact correlator $U(\phi)$, eq.(\ref{1gdp2}), is replaced by its truncated expansion, eq.(\ref{1gdp17}))
could be just fatal for the physical meaning of the obtained results.

\begin{figure}[h]
\begin{center}
   \includegraphics[width=17.0cm]{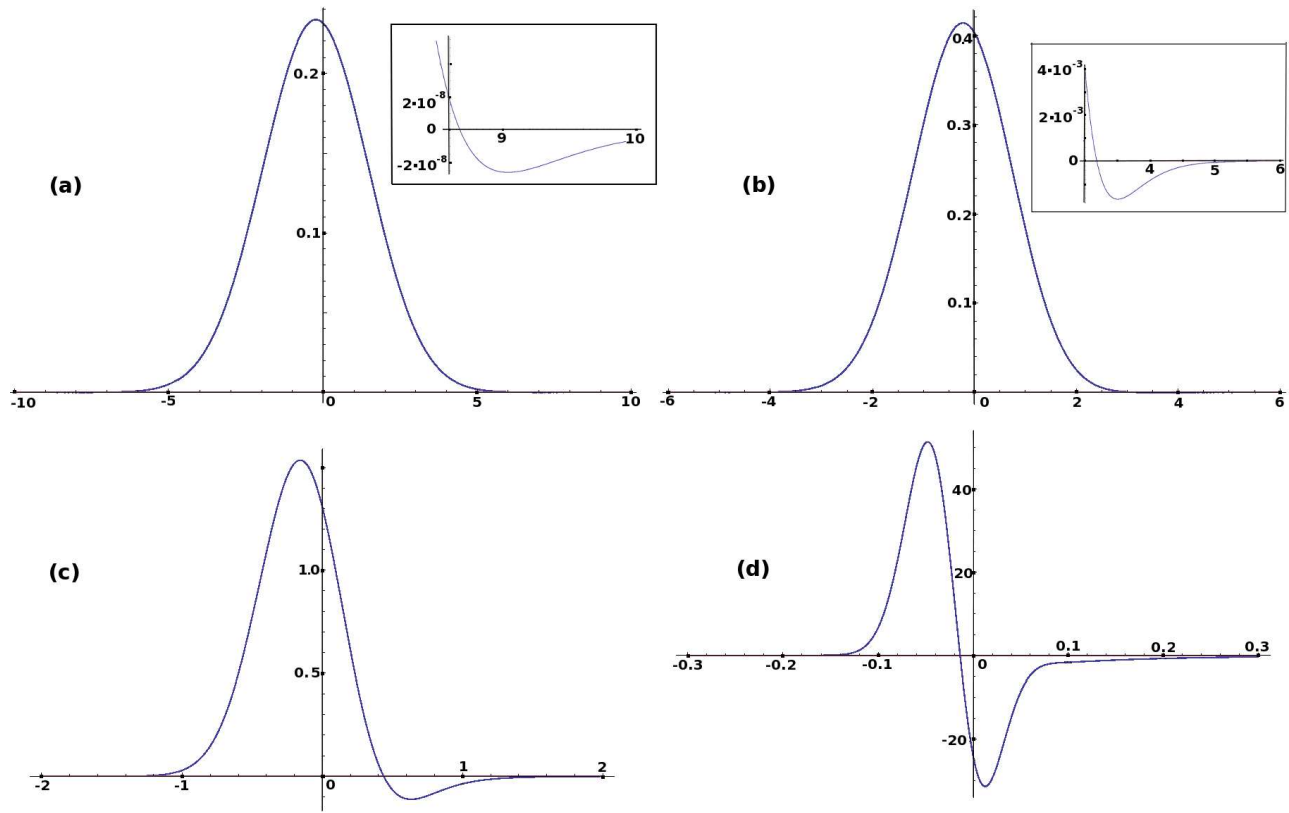}
\caption[]{ Free energy "distribution function" $\tilde{{\cal P}}^{*}_{L}(f)$ 
as it given by eq.(\ref{3gdp7})
for: (a) $\epsilon = 1.5 $, (b) $\epsilon = 0.5 $, (c) $\epsilon =0.05 $, (d) $\epsilon = 0.0005 $,}
\end{center}
\label{figure3}
\end{figure}

\section{Discussion}

The standard program of the replica method is formulated as follows: first, for an arbitrary positive integer
$n$ we have to calculate the disorder average of the $n$-th power of the partition function,
$\overline{Z^{n}} \equiv Z(n)$ which is expected to be an analytic function of the replica parameter $n$;
second, we have to perform an analytic continuation of this function from integer for arbitrary real 
or complex values of $n$; and third, we have to take the limit $n\to 0$ (if we are interested in the 
average free energy only) or we have to perform an integration over complex $n$ (if we are deriving the
free energy distribution function). This third step is usually accompanied by taking the thermodynamic limit,
which assumes that the system size $L$ is taken to infinity. 
The prescription of the replica method indicates that the two limits, $n\to 0$ and $L\to\infty$, 
has to be taken simultaneously such that the product
$n L^{\omega}$ (where an exponent $\omega$ defines the scaling of the free energy with the system size)
is kept finite.

In fact, the whole experience of the replica calculations in 
disordered systems shows that except for trivial cases this program,
as it is formulated above, is never followed! The typical illustration of the
replicas {\it realpolitik}  is provided by the studies in 
the mean-field spin glasses \cite{SK}.
First of all, since the system is sufficiently complicated the computation
of the replica partition function $Z(n)$  
can be done here only in the saddle-point approximation and not exactly.
This makes further analytic continuation to non-integer $n$ somewhat doubtful 
because the neglected terms which are small at {\it integer} $n$ and large system size 
could become essential in the limit $n\to 0$. Moreover, it turns out that
at large $n$ the replica partition function growth as $\sim\exp(n^{2})$ which means
that its analytic continuation to non-integer $n$ is ambiguous. If nevertheless
we would just plainly take the limit $n\to 0$
in the obtained expression for $Z(n)$ , we would get the so called replica symmetric (RS) solution
which at low temperatures is unphysical since it reveals negative entropy and many other
bad things. In view of the remarks made above, of course, this is not surprising. 
The strategy, which is called the replica symmetry breaking (RSB) scheme \cite{RSB-general},
and  which is generally believed to provide correct results, is essentially
different. In this scheme, all the above three steps, (computing $Z(n)$, analytic continuation in 
$n$ and the limits $n\to 0$ and $L\to\infty$) are performed simultaneously!
Similar (although slightly simplified) scheme works perfectly well also in the case
of the Random Energy Model of spin glasses \cite{REM}. For both models the replica results 
are confirmed by independent mathematically rigorous calculations \cite{guerra}.
There are many other systems for which the results of the replica calcualtions, although can not be
confirmed rigorously, are generally accepted to be correct \cite{replicas}.

To understand what is going on sometimes it is useful to consider an example of very simple system.
Replica calculations for the model studied in this paper does not involve any kind of the RSB "magic".
The model is so simple that initially one gets an illusion that every step of the calculations 
could be under good control. For instance, unlike the above example, the replica partition 
function here can be computed exactly for any (finite) system size, eq.(\ref{2gdp27}).
Nevertheless, proceeding with further steps of the replica method program one finds
that either the attempt should be aborted, or one has to start cheating again. 
Indeed, the result, eq.(\ref{2gdp27}), for the replica partition function turns out to be
valid only for {\it finite} number of the integer points: $n \; < \; n_{c} = [\pi^{2}/(4\beta u L^{2})]$,
eq.(\ref{2gdp24}), since at $n > n_{c}$ the quantity $Z(n)$ is not defined (it is formally divergent).
Moreover, for sufficiently large system size, $L > L_{c}(n) = \pi/(2\sqrt{\beta u})$, the replica partition
function $Z(n)$ is not defined for {\it all} positive integer $n$ (including $n=1$). In other words,
in this situation the replica partition function of the considered system, defined as
$\overline{Z^{n}}$ (where $n$ is a positive integer) simply does not exist!
But even if $L < L_{c}(n)$, so that $Z(n)$ is still defined at finite number of the integer points, 
it is evident that its analytic continuation 
for non-integer values of $n$ is completely ambiguous. 
If nevertheless, one neglects all the above observations and accept the result, eq.(\ref{2gdp27}),
as valid for all {\it real} $n$ in the interval $0 < n < n_{c}$, then everything becomes just fine.
The analytic continuation of the function, eq.(\ref{2gdp27}), from the finite interval $0 < n < n_{c}$
to the the complex half-plain, Re$\{n\} < n_{c}$ is unambiguous, 
and in this way one obtains beautiful and {\it correct} results for the free energy distribution 
function (Figures 1 and 2). In this particular case one can be sure that obtained results 
are indeed correct as for the system
under consideration  the quantity $\overline{Z^{n}}$ can be computed directly
for any {\it real} $n \in [0, n_{c}]$ \cite{gaussian}.

All the experience of last decades convincingly demonstrate that
with a few exceptions  the replica method does give correct
results, and this can not be explained by simple coincidences. We know very well {\it how} it works, 
and we do know that the replica calculations inevitably involves cheating. 
The question is then, {\it why} it works?



\begin{thebibliography}{99}

\bibitem{zirnbauer1} J.J.M.Verbaarschot and M.R.Zirnbauer,
         {\it Critic of the replica trick}, 
         J.Phys A: Math. Gen. {\bf 17}, 1093 (1985)


\bibitem{zirnbauer2} M.R.Zirnbauer,
         {\it Another critic of the replica trick}, 
         arXiv: cond-mat/9903338 (1999)

\bibitem{kardar1} M.Kardar,
         {\it Replica Bethe ansatz studies of two-dimensional interfaces with quenched random impurities},
         Nucl. Phys. {\bf B 290}, 582 (1987).

\bibitem{kardar2} E.Medina and M.Kardar,
         {\it Nonuniversality and analytical continuation in moments of directed polymers on hierarchical lattices},
         J. Stat. Phys. {\bf 71}, 967 (1993).

\bibitem{dirpoly} V.S.Dotsenko, L.B.Ioffe, V.B.Geshkenbein, S.E.Korshunov  and G.Blatter, 
         {\it Joint free energy distribution in the random directed polymer problem},
         Phys. Rev. Lett. {\bf 100}, 050601 (2008)

\bibitem{REM} B. Derrida, 
         {\it Random-energy model: An exactly solvable model of disordered systems},
         Phys. Rev. B {\bf 24}, 2613 (1981).


\bibitem{SK} D. Sherrington and S. Kirkpatrick,
        {\it Solvable Model of a Spin-Glass},
        Phys. Rev. Lett. {\bf 35}, 1792 (1975).

\bibitem{RSB-general} M. Mezard, G. Parisi and M. A. Virasoro,
        {\it Spin Glass Theory and Beyond}, 
        World Scientific (Singapore) 1987 

\bibitem{guerra} Adriano Barra, Aldo Di Biasio and Francesco Guerra,
        {\it Replica symmetry breaking in mean field spin glasses through Hamilton-Jacobi technique},
        arXiv: 1003.5226 (2010)

\bibitem{Dotsenko} V.Dotsenko,
    EPL, {\bf 90},20003 (2010); 
   J.Stat.Mech. P07010 (2010)  

\bibitem{LeDoussal} P.Calabrese, P. Le Doussal and A.Rosso
      EPL, {\bf 90},20002 (2010).
    
\bibitem{kanzieper1} E.Kanzieper,
        {\it Replica field theories, Painlev\'e transcendents and exact correlation functions},
        Phys. Rev. Lett. {\bf 89}, 250201 (2002).

\bibitem{splittorff} K.Splittorff and J.J.M.Verbaarschot,
        {\it Replica limit of the Toda lattice equation},
        Phys. Rev. Lett. {\bf 90}, 041601 (2003).

\bibitem{kanzieper2} V.Al.Osipov and E.Kanzieper,
        {\it Are bosonic replicas faulty?},
        Phys. Rev. Lett. {\bf 99}, 050602 (2007).

\bibitem{kanzieper3} E.Kanzieper,
        {\it Replica approach in random matrix theory},
        arXiv: cond-mat/0903.3198v1 (2009).

\bibitem{gaussian} V.S.Dotsenko,  V.B.Geshkenbein, D.A.Gorokhov  and G.Blatter, 
    {\it Free energy distribution function for randomly forced directed polymer}
    arXiv:1007.0852  (2010).

\bibitem{hh_zhang_95} T.\ Halpin-Healy and Y-C.\ Zhang, 
    {\it Kinetic roughening phenomena, stochastic growth, directed polymers and all that. 
    Aspects of multidisciplinary statistical mechanics}, 
    Phys.\ Rep.\ {\bf 254}, 215 (1995).


\bibitem{numer1} D.A.\ Huse and C.L.\ Henley,
    {\it Pinning and Roughening of Domain Walls in Ising Systems Due to Random Impurities},
    Phys.\ Rev.\ Lett. {\bf 54}, 2708 (1985);

\bibitem{numer2} M.\ Kardar and Y-C.\ Zhang,
    {\it Scaling of Directed Polymers in Random Media},
    Phys.\ Rev.\ Lett.\  {\bf 58}, 2087 (1987).

\bibitem{burgers} D.A.\ Huse, C.L.\ Henley, and D.S.\ Fisher,
    Phys.\ Rev.\ Lett.\ {\bf 55}, 2924 (1985).


\bibitem{gorokhov-blatter} D.A.\ Gorokhov and G.\ Blatter, 
    {\it Exact Free Energy Distribution Function of a Randomly Forced Directed Polymer},
    Phys.\ Rev.\ Lett. {\bf 82}, 2705 (1999).

\bibitem{denis_99} D.A.\ Gorokhov and G.\ Blatter, 
    {\it Singularities of the renormalization-group flow for random elastic manifolds},
    Phys.\ Rev.\ B {\bf 59}, 32 (1999).


\bibitem{replicas} V.S. Dotsenko, 
    {\it Introduction to the Replica Theory of Disordered Statistical Systems},
    Cambridge University Press, (2001).

\end{thebibliography}
\end{document}